\documentclass[preprint,aps]{revtex4-2}

\usepackage[utf8]{inputenc} 
\usepackage{graphicx}
\usepackage{bm}
\usepackage{nicefrac}
\usepackage{array} 
\usepackage{upgreek}
\usepackage{amsmath}
\usepackage{makecell} 
\usepackage{booktabs}
\usepackage{multirow} 
\usepackage{titlesec}
\usepackage{placeins} 
\usepackage{hyperref}
\usepackage{cleveref}
\usepackage{tabularx}

\crefname{figure}{Fig.}{Figs.}
\crefname{equation}{Eq.}{Eqs.}

\begin{document}

\raggedbottom

\titleformat{\section}
  {\normalfont\large\bfseries}{\thesection}{1em}{}
\titleformat{\subsection}
  {\normalfont\normalsize\bfseries}{\thesubsection}{1em}{}
\titleformat{\subsubsection}
  {\normalfont\normalsize\itshape}{\thesubsubsection}{1em}{}

\renewcommand\thesection{\arabic{section}}
\renewcommand\thesubsection{\thesection.\arabic{subsection}}
\renewcommand\thesubsubsection{\thesubsection.\arabic{subsubsection}}

\title{Suppression of $p$-Wave Altermagnetism by Localized $4f$ Electrons in CeNiAsO}
\author{Jiuxiang Zhang$^{1,2,\sharp}$, Yueyang Sun$^{1,2,3,\sharp}$, Honglin Zhou$^{1,2,\sharp}$, Jumin Shi$^{1,2,\sharp}$, Di Wu$^{1,2}$, Hongze Gu$^{1,2}$, Wenjin Mao$^{1,2}$, Hengrui Dong$^{1,2}$, Yu Xu$^{1,2}$, Yinghao Li$^{1,2}$, Ziling Cao$^{1,2}$, Taimin Miao$^{1,2}$, Bo Liang$^{1,2}$, Neng Cai$^{1,2}$, Wenpei Zhu$^{1,2}$, Mingkai Xu$^{1,2}$, Jiaqi Chen$^{1,2}$, Chunhong Deng$^{1,2}$, Bo Liu$^{1,2}$, Xun Ma$^{4}$, Zhengtai Liu$^{5,6}$, Mao Ye$^{5}$, Shenjin Zhang$^{7}$, Zhimin Wang$^{7}$, Fengfeng Zhang$^{7}$, Feng Yang$^{7}$, Qinjun Peng$^{7}$, Zuyan Xu$^{7}$, Guodong Liu$^{1,2,8}$, Xintong Li$^{1,2,8}$, Hanqing Mao$^{1,2,8}$, Shiliang Li$^{1,2,*}$, Hongming Weng$^{1,2,3,*}$, Lin Zhao$^{1,2,8,*}$ and X. J. Zhou$^{1,2,8,*}$}
\affiliation{
\\$^{1}$Beijing National Laboratory for Condensed Matter Physics, Institute of Physics, Chinese Academy of Sciences, Beijing 100190, China
\\$^{2}$University of Chinese Academy of Sciences, Beijing 100049, China
\\$^{3}$Condensed Matter Physics Data Center of Chinese Academy of Sciences, Beijing 100190, China
\\$^{4}$School of Nuclear Science and Technology, University of Science and Technology of China, Hefei 230026, China
\\$^{5}$Shanghai Synchrotron Radiation Facility, Shanghai Advanced Research Institute, Chinese Academy of Sciences, Shanghai 201210, China
\\$^{6}$National Key Laboratory of Materials for Integrated Circuits, Shanghai Institute of Microsystem and Information Technology, Chinese Academy of Sciences, Shanghai 200050, China
\\$^{7}$Technical Institute of Physics and Chemistry, Chinese Academy of Sciences, Beijing, China
\\$^{8}$Songshan Lake Materials Laboratory, Dongguan 523808, China
\\$^{\sharp}$These authors contributed equally to the present work.
\\$^{*}$Corresponding authors: slli@iphy.ac.cn, hmweng@iphy.ac.cn, LZhao@iphy.ac.cn, XJZhou@iphy.ac.cn}

\date{\today}

\maketitle

{\bf Altermagnetism, characterized by momentum-dependent spin splitting and zero net magnetization, has so far been explored mainly in weakly or moderately correlated $d$-electron systems. How symmetry-allowed altermagnetic band splitting manifests in heavy-fermion materials, where magnetic exchange competes with Kondo correlations, remains unclear. Here we use high-resolution angle-resolved photoemission spectroscopy (ARPES) to investigate CeNiAsO, a Kondo-lattice system that was predicted to be a candidate for $p$-wave altermagnetism. Fermi surface mapping and polarization-dependent ARPES show that the experimentally observed itinerant bands are mainly derived from Ni $3d$ orbitals, while resonant photoemission reveals that the Ce $4f$ states remain predominantly localized with residual $c$-$f$ hybridization. Ultra-low-temperature measurements reveal no resolvable near-Fermi-level $p$-wave-like exchange splitting on the Ni $3d$-derived conduction bands across the successive antiferromagnetic transitions. These experimental observations cannot be captured by an itinerant-$4f$ band-structure description, which predicts a sizable $p$-wave splitting in the itinerant bands. When the localized Ce $4f$ character is incorporated, our band structure calculations indicate that the itinerant Ce $4f$ band weight is shifted away from the Fermi level and the $p$-wave-like splitting on the Ni $3d$-derived bands is reduced to the few-meV scale. These results establish CeNiAsO as a strongly correlated $f$-electron setting in which the magnetic symmetry allows $p$-wave-like band splitting, but localized $4f$ electrons strongly suppress its observable itinerant single-particle signature.}

\section*{Introduction}

Altermagnetism has recently broadened the classification of long-range magnetic order beyond the conventional ferromagnetic and antiferromagnetic dichotomy\cite{smejkalConventional2022, smejkalEmerging2022, mazinEditorial2022}. In an altermagnet, opposite-spin sublattices are related by a real-space rotation or mirror operation rather than by a simple translation or inversion in the spin-space-group description\cite{jungwirthSymmetry2026,liuSymmetry2026}. This symmetry relation allows momentum-dependent, non-relativistic spin splitting of electronic bands even in the absence of spin-orbit coupling (SOC). The combination of spin-split bands and vanishing net macroscopic magnetization makes this class of compensated magnets attractive for spintronics and quantum information applications\cite{jungwirthAltermagnetic2025,liSpinPolarized2026}.

Experimentally investigated altermagnetic candidates, including MnTe\cite{krempaskyAltermagnetic2024, leeBroken2024} and MnTe$_2$\cite{yuanPrediction2021,zhuObservation2024}, CrSb\cite{reimersDirect2024,santhoshAltermagnetic2025,yangThreedimensional2025}, RuO$_2$\cite{fedchenkoObservation2024,liuAbsence2024}, KV$_2$Se$_2$O\cite{jiangmetallic2025} and RbV$_2$Te$_2$O\cite{zhangCrystalsymmetrypaired2025}, are mostly weakly or moderately correlated $d$-electron systems. Closely related odd-parity $p$-wave magnets have also been proposed in non-collinear magnetic materials such as NiI$_2$\cite{songElectrical2025} and Gd$_3$(Ru$_{1-\delta}$Rh$_{\delta}$)$_4$Al$_{12}$\cite{yamadametallic2025}. In these systems, the spin-split band structure is largely understood within a single-particle band theory governed by crystal and spin symmetries. Whether the same picture survives in strongly correlated environments, particularly $f$-electron heavy-fermion systems, remains unclear. In such systems, the low-energy electronic structure is governed by the Doniach competition \cite{doniachKondo1977}: the Ruderman–Kittel–Kasuya–Yosida (RKKY) interaction promotes long-range magnetic order, whereas Kondo correlations screen local moments and generate low-energy $f$-derived quasiparticle spectral weight through $c$-$f$ hybridization \cite{siHeavy2010}. How this many-body competition reshapes symmetry-allowed altermagnetic band splitting is an open question.

CeNiAsO provides a concrete setting to address this question. It is an antiferromagnetic Kondo lattice with successive antiferromagnetic transitions at $T_{N1}\approx 9~\mathrm{K}$ and $T_{N2}\approx 6\text{--}7~\mathrm{K}$\cite{luoCeNiAsO2011, luoHeavyfermion2014, wuIncommensurate2019}. For its low-temperature noncollinear antiferromagnetic phase below $T_{N2}$, density functional theory (DFT) calculations predict an odd-parity exchange splitting on the itinerant conduction bands, making CeNiAsO a candidate for metallic $p$-wave magnetism\cite{hellenesPwave2024, chakrabortyHighly2025}. Here we refer to this odd-parity compensated magnetic state more broadly as $p$-wave altermagnetism, and focus on its expected single-particle signature: a momentum-dependent $p$-wave exchange splitting of the itinerant bands. Recent transport measurements have reported a large field-trained in-plane resistivity anisotropy below $T_{N2}$\cite{zhouAnisotropic2025}, consistent with broken tetragonal symmetry in the magnetically ordered state. However, transport anisotropy does not directly determine whether the itinerant bands carry the anticipated single-particle exchange splitting. The microscopic manifestation of the predicted $p$-wave splitting in the electronic structure of CeNiAsO therefore remains to be clarified.

Here we use high-resolution laser-based angle-resolved photoemission spectroscopy (ARPES) to investigate the electronic structure of CeNiAsO across its magnetic transitions. We find no resolvable $p$-wave-like exchange splitting on the itinerant Ni $3d$-derived conduction bands. By combining Fermi surface mapping, orbital-resolved ARPES, resonant photoemission, and correlation-dependent DFT calculations, we show that the measured Fermi surface is dominated by Ni $3d$-derived itinerant bands, whereas the Ce $4f$ states remain predominantly localized with residual $c$-$f$ hybridization. The large splitting expected in an itinerant-$4f$ picture is therefore not the experimentally relevant band splitting. Instead, $4f$-electron localization suppresses the transfer of the symmetry-allowed $p$-wave exchange splitting to the measured Ni $3d$ bands, reducing the observable $p$-wave-like splitting to the few-meV scale. Our results identify CeNiAsO as a strongly correlated $f$-electron setting in which the symmetry-allowed $p$-wave-like itinerant band splitting is strongly suppressed.

\section*{Results}
CeNiAsO crystallizes in a tetragonal ZrCuSiAs-type structure (space group $P4/nmm$), where the Ce $4f$ moments form a coplanar noncollinear antiferromagnetic order below $T_{N2}$ (Fig. 1a). The corresponding Brillouin zone (BZ) and the (001) surface-projected BZ are shown in Fig. 1b. The high quality of our CeNiAsO single crystals is confirmed by sharp Laue diffraction patterns on the (001) surfaces (see Supplementary Fig. S1b). Macroscopic magnetization measurements reveal two successive magnetic phase transitions at $T_{N1}\approx8.7$\,K and $T_{N2}\approx6.1$\,K for our CeNiAsO samples (Fig. 1c). The phase below $T_{N2}$ is of particular interest because a metallic $p$-wave magnetic state has been proposed to be realized in the coplanar noncollinear magnetic structure\cite{hellenesPwave2024, chakrabortyHighly2025}.

We first determine the Fermi surface topology by combining synchrotron- and laser-based ARPES measurements. Synchrotron measurements at $h\nu=75$\,eV (Fig. 1d) cover an extended momentum region and resolve the bulk-derived Fermi surface contours (Fig. 1d), whereas the high-resolution laser ARPES at $h\nu=6.994$\,eV provides a detailed view of the near-$E_F$ states around $\bar{\Gamma}$, where both surface and bulk contributions are visible (Fig. 1e). By tracking the constant energy contours (Fig. 1d and Fig. 1e) and analyzing the related band structures (Fig. 2), we extract the measured Fermi surface of CeNiAsO as shown in Fig. 1f. The innermost near-$\bar{\Gamma}$ features are assigned to surface states (denoted as SS, thick gray line in Fig. 1f), as supported by the CeO-terminated slab calculation shown in Fig. 2g. The CeO termination of the measured surface is identified from the termination-dependent Fermi surface topology and work-function difference presented in Supplementary Fig. S2. The remaining Fermi surface sheets are assigned to bulk itinerant states: three electron-like pockets ($\alpha$, $\beta_1$ and $\beta_2$) are centered around the $\bar{\mathrm{M}}$ point, while a hole-like pocket ($\gamma$) is centered around the $\bar{\mathrm{X}}$ point. 

We then compare the measured Fermi surface topology with first-principles calculations under different treatments of the Ce $4f$ electrons. When the Ce $4f$ states are included as valence electrons, the corresponding Fermi surface (Fig. 1g), similar to previous calculations\cite{hellenesPwave2024}, fails to reproduce the measured Fermi surface topology. In particular, the two Fermi pockets around $\bar{\Gamma}$ in Fig. 1g, which are derived mainly from Ce $4f$ orbitals in the calculation, are absent in the measured Fermi surface in Fig. 1f. By contrast, when the Ce $4f$ states are treated as core-like and excluded from the valence electrons, the calculated Fermi surface (Fig. 1h) closely resembles the measured bulk Fermi surface (Fig. 1f). In this case, all the Fermi surface sheets in the calculation are derived from Ni $3d$ orbitals. These results indicate that the low-energy electronic states are mainly derived from Ni $3d$ orbitals, while the Ce $4f$ electrons are largely localized. 

Figures 2a-c show the band structures of CeNiAsO measured by synchrotron-based ARPES with a photon energy of $h\nu=75$\,eV. Along the high-symmetry $\bar{\Gamma}$--$\bar{\mathrm{X}}$, $\bar{\Gamma}$--$\bar{\mathrm{M}}$ and $\bar{\mathrm{X}}$--$\bar{\mathrm{M}}$ directions, the dominant dispersive features can be assigned to the bulk $\alpha$, $\beta_1$, $\beta_2$ and $\gamma$ Fermi surface identified in Fig. 1. The observed band structures are well reproduced by the nonmagnetic DFT calculations ($4f$ in core) along the $\mathrm{M}$--$\Gamma$--$\mathrm{X}$--$\mathrm{M}$ path (Fig. 2f), confirming that the itinerant electronic structure is mainly derived from Ni $3d$ states. In addition, surface states are clearly observed around $\bar{\Gamma}$ and $\bar{\mathrm{X}}$, labeled as SS in Fig. 2a-c. The SS surface states are more clearly observed in laser ARPES measurements (Fig. 2d,e) and consist of two branches. Comparison with slab calculations for the CeO-terminated surface (Fig. 2g) supports the assignment of these branches to the surface states. A weak nearly nondispersive feature is also clearly observed at a binding energy of $\sim270$\,meV, which can be assigned to the Ce $4f^1_{7/2}$ state as marked in Fig. 2c.

The orbital character of the observed bands can be further revealed by polarization-dependent ARPES measurements (Fig. 2d,e) and band structure calculations (Fig. 2f). By switching the incident light polarization ($s$- and $p$-polarizations), initial states with different parities with respect to the experimental mirror plane are selectively enhanced. Along the $\bar{\Gamma}$--$\bar{\mathrm{X}}$ direction (Fig. 2d), the $\gamma$ band is observed under $s$-polarization geometry (left panel of Fig. 2d) but is strongly suppressed under $p$-polarization geometry (right panel of Fig. 2d). The matrix element analysis restricts the possible Ni $3d$ orbital components of the $\gamma$ band to $d_{yz}$ and $d_{xy}$. This restriction is further supported by the orbital-resolved Ni $3d$ band projections (Fig. 2f and Supplementary Fig. S4), where the $\gamma$ band along $\Gamma$--$\mathrm{X}$ consists mainly of Ni $d_{yz}$ and $d_{xy}$ orbitals. Along the $\bar{\Gamma}$--$\bar{\mathrm{M}}$ direction (Fig. 2e), the $\alpha$ band shows the same $s$-polarized selection behavior, being visible in the $s$-polarization geometry and suppressed in the $p$-polarization geometry. Together with the calculated orbital projections, this identifies the $\alpha$ band as mainly derived from Ni $d_{x^2-y^2}$, $d_{xz}$, and $d_{yz}$ components. The orbital components of all the bulk bands along different high symmetry directions are summarized in the orbital-resolved band projections in Fig. 2f and Supplementary Fig. S4.

To further probe the role of Ce $4f$ orbitals in the low-energy electronic structure, we performed resonant ARPES across the Ce $4d\rightarrow4f$ resonance edge. Figures 3a and 3b compare the band structures of CeNiAsO measured along $\bar{\Gamma}$--$\bar{\mathrm{X}}$ off resonance ($h\nu=105$\,eV) and on resonance ($h\nu=120$\,eV), respectively. At resonance, the spectra show a pronounced enhancement of nearly nondispersive Ce $4f$-derived spectral weight. The EDC comparison at $\bar{\Gamma}$ resolves the characteristic Ce $4f$ multiplet structure, including the near-$E_F$ $4f^1_{5/2}$ resonance and the spin-orbit-split $4f^1_{7/2}$ satellite at $E-E_F\approx-0.28$\,eV (Fig. 3c). In Ce-based resonant-photoemission studies, the near-$E_F$ $4f^1_{5/2}$ resonance is commonly associated with the Kondo-resonance channel, reflecting finite mixing between localized Ce $4f$ states and itinerant conduction electrons\cite{fujimoriDirect2006,patilARPES2016, chenDirect2017}. The deeper $f^0$ final-state feature associated with removal of a Ce $4f$ electron typically appears at $1.5$--$2.5$\,eV below $E_F$, outside our experimentally accessible window extending from $E_F$ to approximately 1.2\,eV below $E_F$\cite{fujimoriDirect2006,patilARPES2016}. These resonantly enhanced but weakly dispersive features therefore support predominantly localized Ce $4f$ character with residual many-body $c$-$f$ hybridization, rather than dispersive itinerant Ce $4f$ bands crossing the Fermi level. In particular, although the Ce $4d\rightarrow4f$ resonance strongly enhances the Ce $4f$ photoemission cross section, we do not observe the dispersive itinerant $4f$-derived bands predicted in the uncorrelated calculations\cite{hellenesPwave2024, chakrabortyHighly2025}.  

We then assess how the treatment of the Ce $4f$ electrons affects the calculated $p$-wave exchange splitting in CeNiAsO. In the $U_{\mathrm{eff}}=0$ calculation, the Ce $4f$ states are treated as itinerant valence states and contribute substantial spectral weight near the Fermi level (Fig. 3e)\cite{hellenesPwave2024, chakrabortyHighly2025}. The corresponding spin-polarized band structures are shown in Fig. 3f. As a result, the calculated spin-projected Fermi surface for the proposed $p$-wave magnetic configuration (Fig. 3g) contains additional Ce $4f$-derived pockets that are absent in the ARPES data. The large $p$-wave exchange splitting associated with these itinerant $4f$-derived pockets ($\sim20$\,meV) is therefore not directly relevant to the experimentally observed Fermi surface. By examining the calculated constant energy contours at different binding energies (see Supplementary Fig. S5), we find that the contour at $E-E_F=-0.25$\,eV is qualitatively consistent with the observed Fermi surface in Fig. 1. After rigidly shifting the Fermi level to $E_F^\ast=E_F-0.25$\,eV (as marked by dashed green lines in Fig. 3e and 3f), the corresponding spin-projected Fermi surface for the proposed $p$-wave magnetic configuration is presented in Fig. 3h. In this case, the electronic states near the new Fermi level are dominated by Ni $3d$ orbitals. As a result, the calculated $p$-wave-like exchange splitting on the Ni $3d$-derived bands is $\sim10$\,meV (Fig. 3f).

By contrast, a more appropriate treatment for a Ce-based Kondo-lattice system is to include a sizable on-site Coulomb interaction on the Ce $4f$ orbitals. As shown in Supplementary Fig. S7a, we take $U_{\mathrm{eff}}=6$\,eV as a representative predominantly localized-$4f$ limit because it removes the uncorrelated dispersive Ce $4f$ bands from the $\sim1.2$\,eV energy window (Fig. 3a and 3b). For $U_{\mathrm{eff}}=6$\,eV, the itinerant Ce $4f$ band weight is shifted to higher binding energies, substantially reducing the $4f$ admixture with the Ni $3d$-derived conduction bands near $E_F$ (Fig. 3i). The resulting calculated Fermi surface for the proposed magnetic configuration is dominated by Ni $3d$-derived pockets, in much closer agreement with the ARPES measurements (Fig. 3k). In this predominantly localized-$4f$ description, the $p$-wave-like exchange splitting on the experimentally relevant Ni $3d$-derived bands is reduced to $\sim5$\,meV (Fig. 3j), remaining at the few-meV scale for $U_{\mathrm{eff}}=3$--$6$\,eV (Supplementary Fig. S7b). The corresponding splitting in momentum space is $\Delta k\sim0.002~\text{\AA}^{-1}$, which is too small to be visually resolved in the plotted band structure (Fig. 3j). The calculated splitting values for the Ce $4f$- and Ni $3d$-derived bands in the itinerant-$4f$ reference, the rigidly shifted Ni $3d$ reference, and the localized $4f$ DFT+$U$ case are summarized in Fig. 3d.

To directly examine whether band splitting develops in the antiferromagnetic states of CeNiAsO, we performed temperature-dependent ARPES measurements across the magnetic transitions. Figures 4a,b show band structures measured at 9\,K and 6.2\,K along $\bar{\Gamma}$--$\bar{\mathrm{X}}$ (Cut 1 in Fig. 4d) by synchrotron-based ARPES at a photon energy of $75~\mathrm{eV}$, and the corresponding Fermi-level MDCs are compared in Fig. 4c. Along this cut, the $\gamma$ band is observed together with the SS surface states. Upon cooling through $T_{N1}\approx8.7~\mathrm{K}$, neither an obvious band splitting nor band folding is detected for the $\gamma$ band.

To further search for subtle changes in the Ni $3d$-derived bands, we carried out high-resolution laser ARPES measurements at $h\nu=6.994~\mathrm{eV}$. Figures 4e,f show band structures measured at 15\,K and 1.6\,K along Cut 2, with MDCs at $E_F$ and at $10~\mathrm{meV}$ binding energy shown in Fig. 4g. The $\gamma$ band exhibits little change between 15\,K and 1.6\,K, with the latter being well below $T_{N2}\approx6.1~\mathrm{K}$. Similarly, Figs. 4h,i show band structures measured along Cut 3, and the corresponding MDCs are shown in Fig. 4j. The $\alpha$ band along this cut also remains essentially unchanged across the same temperature range. These results show that no resolvable $p$-wave-like splitting, nor any band folding, develops on the observed Ni $3d$-derived bands across the magnetic transitions. The calculated momentum separation in the localized-$4f$ case, $\Delta k\sim0.002~\text{\AA}^{-1}$, is comparable to or smaller than the practical resolving capability set by the experimental momentum resolution and the finite MDC linewidths. The absence of resolvable splitting in Fig. 4 is therefore consistent with the localized-$4f$ picture, in which the symmetry-allowed exchange splitting is reduced to a scale that is not resolved in the present ARPES measurements.

\section*{Discussion and conclusion}

Our results show that the Ce $4f$ electrons in CeNiAsO cannot be treated as ordinary itinerant valence states. When the Ce $4f$ states are included as itinerant bands, the calculated Fermi surface contains additional $4f$-derived pockets that are absent in our ARPES measurements. By contrast, the measured Fermi surface and dispersive low-energy bands are dominated by Ni $3d$ states. Resonant ARPES further supports this assignment: the Ce $4f$ spectral weight appears mainly as weakly dispersive $4f^1_{5/2}$ and $4f^1_{7/2}$ final state features, rather than as dispersive itinerant $4f$ bands crossing the Fermi level. These observations establish that the experimentally relevant itinerant electronic structure is primarily Ni $3d$-derived, while the Ce $4f$ states remain predominantly localized with residual $c$-$f$ hybridization.

This correlated-$4f$ regime requires an explicit treatment of strong on-site Coulomb interaction on the Ce $4f$ orbitals. Introducing $U_{\mathrm{eff}}$ shifts the itinerant Ce $4f$ band weight to higher binding energy and removes the uncorrelated itinerant $4f$ bands from the near-$E_F$ energy window probed by ARPES. This does not exclude the weak near-$E_F$ resonant $4f$ spectral weight observed experimentally, which reflects residual $c$-$f$ hybridization rather than a fully itinerant $4f$ band crossing $E_F$. For this reason, the $U_{\mathrm{eff}}=6$\,eV calculation should be viewed as a representative localized-$4f$ description, rather than as a uniquely fitted parameter. Once the itinerant Ce $4f$ band weight is moved away from the low-energy Ni $3d$-derived conduction bands, the $4f$ admixture with the Ni $3d$ states near $E_F$ is strongly reduced, and the calculated Fermi surface becomes consistent with the measured one.

This localized-$4f$ regime provides a natural microscopic picture for the absence of a large itinerant $p$-wave-like exchange splitting. In a weak-correlation band picture, the symmetry of the magnetic order is expected to be directly reflected in the single-particle band structure. In CeNiAsO, however, the ordered moments are primarily associated with localized Ce $4f$ electrons, whereas the experimentally observed Fermi surface is formed mainly by itinerant Ni $3d$ bands. The transfer of the symmetry-allowed exchange field from the localized $4f$ moments to these conduction bands is therefore limited by the weak residual $c$-$f$ hybridization. As a result, the magnetic symmetry and the observable itinerant band splitting become partially decoupled, leaving only a small $p$-wave-like splitting on the Ni $3d$-derived bands.

This distinction also clarifies the relation between our spectroscopic results and the macroscopic anisotropy reported in transport measurements. A field-trained in-plane resistivity anisotropy below $T_{N2}$ is consistent with broken rotational symmetry in the magnetically ordered state, but it does not by itself require a large single-particle exchange splitting on the itinerant bands. Our ARPES results show that the symmetry-allowed $p$-wave exchange splitting is only weakly transferred to the experimentally observed Ni $3d$ bands. The large splitting obtained in the itinerant-$4f$ reference calculation is mainly associated with $4f$-derived states that do not form the measured Fermi surface. In the localized-$4f$ calculation, the splitting on the Ni $3d$-derived bands is reduced to the few-meV scale, with a momentum separation of only $\Delta k\sim0.002~\text{\AA}^{-1}$. Such a small splitting is at or below the practical resolving capability set by the experimental momentum resolution and the finite MDC linewidths, naturally explaining why no resolvable band splitting is observed across the magnetic transitions. Thus, CeNiAsO can host magnetically driven macroscopic anisotropy while displaying only a strongly suppressed itinerant single-particle band-splitting signature.

In conclusion, our high-resolution ARPES measurements, combined with correlation-dependent first-principles calculations, establish that the Ce $4f$ electrons in CeNiAsO are predominantly localized and cannot be adequately treated as ordinary itinerant valence states. Including a sizable on-site Coulomb interaction shifts the itinerant Ce $4f$ band weight away from the near-$E_F$ Ni $3d$-derived conduction bands and reduces the $p$-wave-like exchange splitting on the experimentally relevant Ni $3d$ states to the few-meV scale. Consistently, ARPES reveals no resolvable $p$-wave-like exchange splitting on the Ni $3d$-derived itinerant bands across the successive magnetic transitions of CeNiAsO. These results identify CeNiAsO as a strongly correlated $f$-electron setting for $p$-wave altermagnetism, where $4f$ localization suppresses the observable single-particle band-splitting signature.

\section*{Methods}
\subsection*{Single crystal growth and characterization}
High-quality CeNiAsO single crystals were grown using the self-flux method. Detailed macroscopic transport properties and additional characterization of CeNiAsO single crystals have been reported previously \cite{zhouAnisotropic2025}. The sample quality for the photoemission experiments is further checked by core-level X-ray photoemission spectroscopy (XPS) and Laue diffraction measurements (Supplementary Fig. S1).

\subsection*{Angle-resolved photoemission spectroscopy (ARPES)}
ARPES measurements were carried out on three complementary setups to achieve comprehensive momentum and energy coverage. High-resolution data were acquired using two laboratory-based systems, both employing a 6.994\,eV vacuum-ultraviolet (VUV) laser as the excitation source\cite{liu2008development, zhou2018new}. On one of these systems, a $-95$\,V bias voltage was applied to the sample to extend the accessible in-plane momentum range (Fig. 1e and Fig. 2d,e); the overall energy resolution was $\sim6$\,meV and the angular resolution of the DA30L analyzer was $\sim0.2^\circ$\cite{miaoExpansion2026}. The second laser-based ARPES setup is configured for ultra-low-temperature operation (Fig. 4e,f and Fig. 4h,i). To enhance the data statistics of the weak signal, the energy resolution was set at $\sim7.5$\,meV. The angular resolution of the DA30 analyzer is $\sim0.2^\circ$, which corresponds to a momentum resolution of $\sim0.003~\text{\AA}^{-1}$ for the photon energy of 6.994\,eV. The lowest sample temperature was 1.6\,K in the laser ARPES measurements. All samples were cleaved \textit{in situ} in an ultra-high vacuum better than $4 \times 10^{-11}$ mbar for the laser ARPES measurements. Wide momentum coverage and photon-energy-dependent ARPES measurements were performed at the BL03U endstation of the Shanghai Synchrotron Radiation Facility (SSRF)\cite{yangHighresolution2021}, equipped with a DA30L analyzer (Fig. 1d, Fig. 2a-c, Fig. 3a,b, Fig. 4a-c and Supplementary Fig. S3). Synchrotron data were taken using a pass energy of 10\,eV and a slit width of 0.3\,mm. Photon-energy-dependent measurements yielded an inner potential $V_0=16$\,eV, placing the 75\,eV data close to the bulk $k_z=0$ ($\Gamma$) plane.

\subsection*{First-principles calculations}
First-principles calculations were performed within density functional theory (DFT) using the projector augmented-wave (PAW) method, as implemented in the Vienna ab initio Simulation Package (VASP). The generalized-gradient approximation (GGA) in the Perdew–Burke–Ernzerhof (PBE) form was adopted for the exchange-correlation functional. The experimental crystal structure of CeNiAsO was used as the structural input. A plane-wave kinetic-energy cutoff of 520\,eV was used for the main DFT+$U$ calculations, while a cutoff of 460\,eV was applied for the plain PBE reference calculations without a Hubbard $U$. The electronic energy convergence criterion was set to $10^{-6}$\,eV for the DFT+$U$ iterations and $10^{-5}$\,eV for the PBE references.

Several reference calculations were performed to distinguish the effects of magnetism, Ce $4f$ localization, and the Ce PAW potential. Plain PBE calculations with the standard Ce PAW potential were first carried out in both nonmagnetic and magnetic configurations, corresponding to the $U_{\mathrm{eff}}=0$ limit with explicit Ce $4f$ valence states retained. The nonmagnetic plain PBE results were used to obtain the constant energy contours at $E-E_F = 0, -0.1, -0.2, -0.25, -0.3$ and $-0.4$\,eV (Supplementary Fig. S5). The contour at $E-E_F=-0.25$\,eV most closely resembles the experimentally observed Ni $3d$-derived Fermi surface topology, and therefore defines the rigid-energy-aligned reference $E_F^\ast=E_F-0.25$\,eV used in Fig. 3. In the magnetic plain PBE reference, a commensurate noncollinear magnetic supercell was used, with spatial symmetries disabled. The Ce moments were constrained to the coplanar low-temperature magnetic configuration by a penalty functional, using the target spin matrix

$$\mathbf{M}_{\mathrm{Ce}} = \begin{pmatrix} 0.66 & -0.44 & 0 \\ -0.66 & -0.44 & 0 \\ -0.66 & 0.44 & 0 \\ 0.66 & 0.44 & 0 \end{pmatrix} \mu_{\mathrm{B}}$$
in Cartesian coordinates. In addition, a nonmagnetic PBE calculation using the Ce$_3$ PAW potential, which suppresses explicit Ce $4f$ valence states, was performed as a clean conduction-band reference (Fig. 1h).

The on-site Coulomb interaction of the Ce $4f$ shell was treated using the rotationally invariant Dudarev DFT+$U$ formalism, with the Hubbard correction applied only to the Ce $4f$ orbitals. Magnetic DFT+$U$ calculations with the standard Ce PAW potential were performed for $U_{\mathrm{eff}} = 1, 2, 3, 4, 5$ and $6$\,eV. The spin-polarized band evolution is shown in Supplementary Fig. S7. In the main text, $U_{\mathrm{eff}} = 6$\,eV is used as a representative localized-$4f$ calculation, comparable to the interaction scale $U-J\approx6.7$\,eV used in previous DMFT calculations of CeNiAsO \cite{luoHeavyfermion2014}. Supplementary Fig. S7 further shows that the Ni $3d$-band splitting changes only weakly within the localized-$4f$ regime.

Spin-orbit coupling (SOC) was included self-consistently in the noncollinear magnetic calculations. The spin expectation values were evaluated in the global spin frame, and the color scales in the spin-polarized band plots represent the expectation value of the out-of-plane spin component ($S_z$). Orbital-resolved band structures and densities of states were obtained by weighting the Kohn-Sham eigenvalues with the corresponding orbital projection coefficients.

To evaluate the fine details of the Fermi surface and generate smooth $S_z$ spin-polarization maps, tight-binding models were constructed. Maximally localized Wannier functions were generated using Wannier90, utilizing spinor Wannier functions for the magnetic and SOC calculations. The projection basis comprised Ce $4f$, Ni $3d$, As $4p$, and O $2p$ orbitals. The validity of the resulting Wannier Hamiltonians was checked by benchmarking the interpolated bands against the initial DFT dispersions within the low-energy window relevant to our ARPES measurements.

Surface states were calculated using a symmetric CeO slab with $\sim21.7~\text{\AA}$ vacuum. Nonmagnetic SOC slab bands were computed along $\Gamma$--$\mathrm{X}$--$\mathrm{M}$--$\Gamma$ using the Ce$_3$ PAW potential. Surface character was evaluated from VASP PROCAR by projecting the wave functions onto the outermost CeO layers.

\section*{Data availability}
All data that support the findings of this study are available from the corresponding authors upon reasonable request.

\section*{Code availability}
The custom codes used for ARPES data analysis and first-principles post-processing are available from the corresponding authors upon reasonable request. 

\newpage


\noindent {\bf Acknowledgements}\\
This work is supported by the National Natural Science Foundation of China (Grant Nos. 12488201 by X.J.Z., 12374066 by L.Z. and 12374154 by X.T.L.), the National Key Research and Development Program of China (Grant Nos. 2021YFA1401800 by X.J.Z., 2022YFA1604200 by L.Z., 2022YFA1403900 by G.D.L. and 2023YFA1406000 by X.T.L.), CAS Superconducting Research Project (Grant No. SCZX-0101), Innovation Program for Quantum Science and Technology (Grant No. 2021ZD0301800 by X.J.Z.), the Youth Innovation Promotion Association of CAS (Grant No. Y2021006 by L.Z.) and the Synergetic Extreme Condition User Facility (SECUF). H.W. acknowledges support from the National Natural Science Foundation of China (Grant No. 12188101), the National Key Research and Development Program of China (Grant No. 2022YFA1403800), and the New Cornerstone Science Foundation through the XPLORER PRIZE. S.L.L. acknowledges support from the National Key Research and Development Program of China (Grant Nos. 2022YFA1403400 and 2021YFA1400400).\\

\noindent {\bf Author contributions}\\
X.J.Z., L.Z., S.L.L. and J.X.Z. proposed and designed the research. J.X.Z. and J.M.S. carried out the ARPES experiments. H.L.Z. and B. Liu grew the single crystals. D.W., H.Z.G., W.J.M., H.R.D., Y.X., Y.H.L., Z.L.C., T.M.M., B. Liang, W.P.Z., N.C., M.K.X., J.Q.C., C.H.D., S.J.Z., Z.M.W., F.F.Z., F.Y., Q.J.P., Z.Y.X., G.D.L., X.T.L., H.Q.M., L.Z. and X.J.Z. contributed to the development and maintenance of the ARPES systems and related software. X.M., Z.T.L. and M.Y. provided the experimental support for the measurements performed on BL03U in SSRF. Y.Y.S. and H.M.W. contributed to theoretical analysis. X.J.Z., L.Z. and J.X.Z. analyzed the data and wrote the paper. All authors participated in discussions and comments on the paper.\\

\noindent {\bf Competing interests}\\
The authors declare no competing interests.

\bibliography{CNAO_project}

\newpage

\begin{figure*}[tbp]
    \centering
    \includegraphics[width=\textwidth]{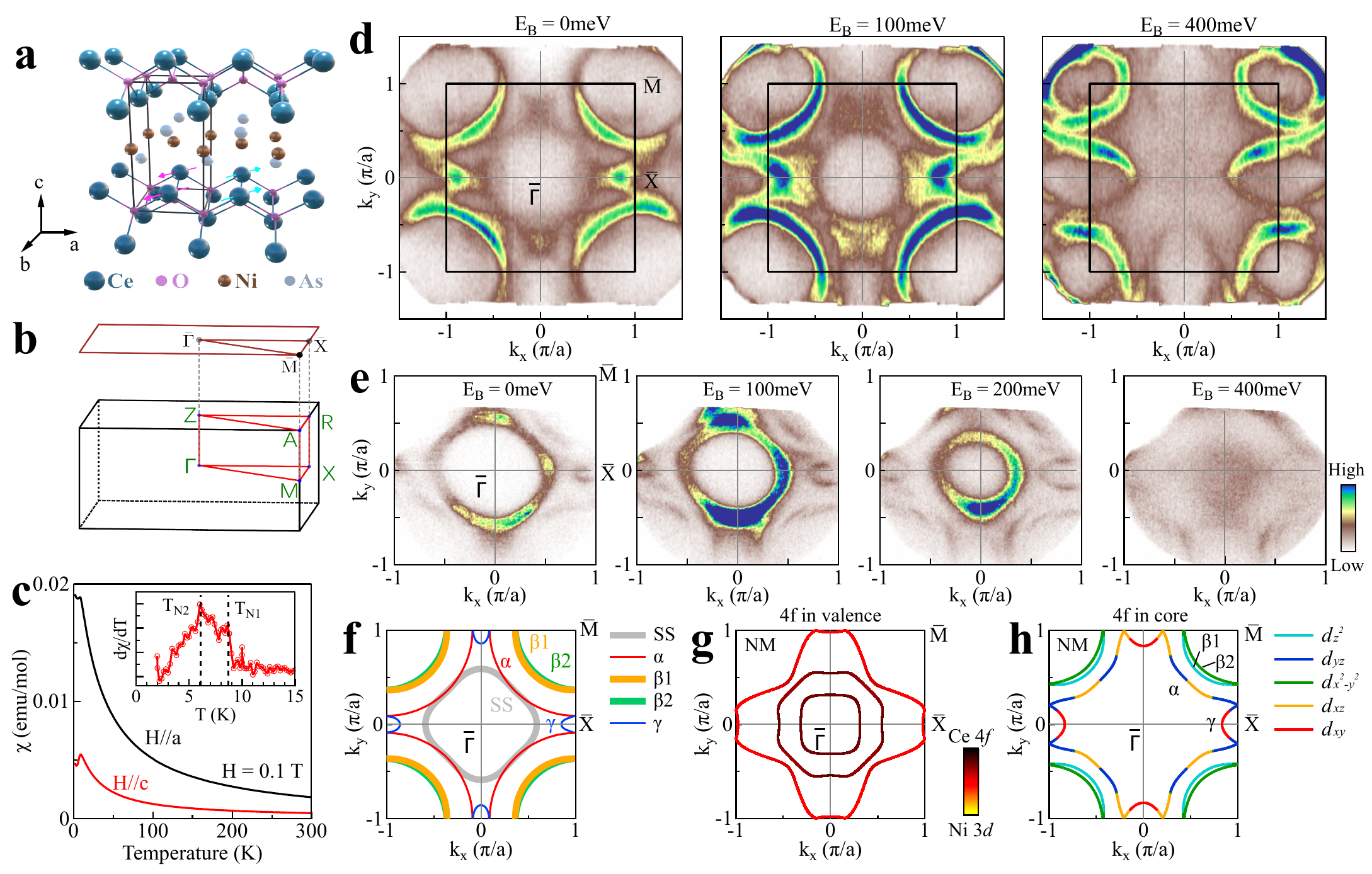}
    \caption{\textbf{Fermi surface determination of CeNiAsO and its comparison with theoretical calculations.} 
    (a) Crystal structure of CeNiAsO. The staggered Ce moments in the antiferromagnetic state below $T_{N2}\approx6.1$\,K are indicated by arrows.
    (b) Three-dimensional bulk Brillouin zone (BZ) and the corresponding (001) surface-projected BZ, with high-symmetry points indicated. 
    (c) Temperature dependence of the magnetic susceptibility $\chi$ measured under $H=0.1$\,T for fields applied parallel and perpendicular to the $c$ axis. The inset shows $d\chi/dT$ with two successive magnetic transitions at $T_{N1}\approx8.7$\,K and $T_{N2}\approx6.1$\,K.
    (d) Constant energy contours measured by synchrotron ARPES at $h\nu=75$\,eV for binding energies $E_B=0$, 100 and 400\,meV. The black rectangle marks the surface-projected first BZ.
    (e) Constant energy contours measured by high-resolution laser ARPES at $h\nu=6.994$\,eV for $E_B=0$, 100, 200 and 400\,meV.
    (f) Experimental Fermi surface extracted from the ARPES measurements in (d) and (e). The inner contour around $\bar{\Gamma}$ is assigned to the surface state (SS), while the $\alpha$, $\beta_1$, $\beta_2$ and $\gamma$ contours denote bulk-derived Fermi surface sheets.
    (g) Fermi surface from nonmagnetic density functional theory (DFT) calculations with Ce $4f$ electrons treated as valence states.
    (h) Fermi surface from nonmagnetic DFT calculations with Ce $4f$ electrons treated as core states.}
    \label{fig:figure1}
\end{figure*}

\begin{figure*}[tbp] 
    \centering
    \includegraphics[width=\textwidth]{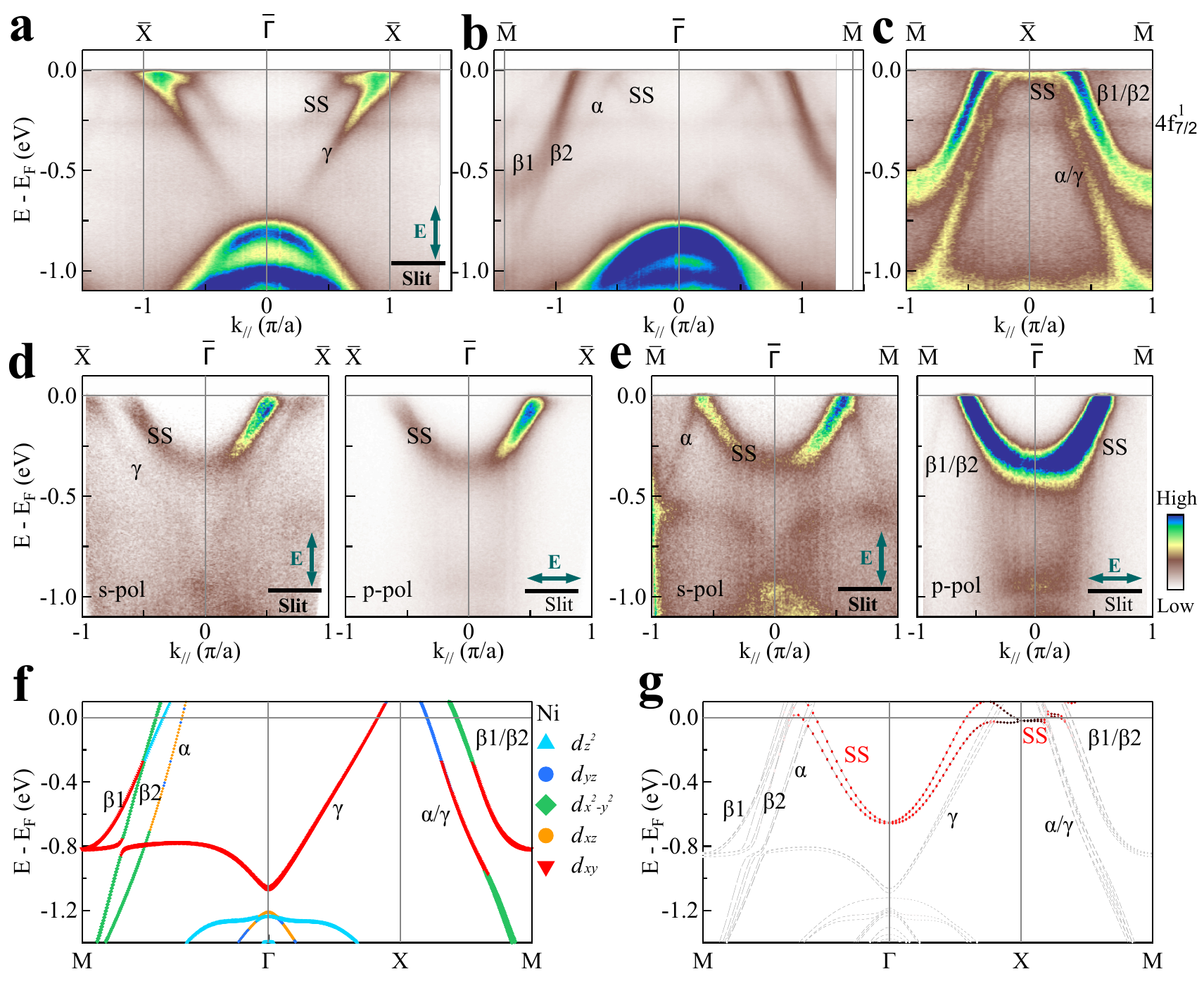}
    \caption{\textbf{Band structures of CeNiAsO measured by synchrotron ARPES and laser ARPES and their comparison with band structure calculations. }
    (a)-(c) Band structures measured at 7\,K by synchrotron ARPES with a photon energy of $h\nu=75$\,eV along $\bar{\mathrm{X}}$--$\bar{\Gamma}$--$\bar{\mathrm{X}}$, $\bar{\mathrm{M}}$--$\bar{\Gamma}$--$\bar{\mathrm{M}}$ and $\bar{\mathrm{M}}$--$\bar{\mathrm{X}}$--$\bar{\mathrm{M}}$, respectively. The main bulk-derived bands $\alpha$, $\beta_{1}$, $\beta_{2}$ and $\gamma$ and the near-$\bar{\Gamma}$ surface states (SS) are marked. A nearly nondispersive feature is observed at a binding energy of 0.27\,eV, which is assigned to the Ce $4f^1_{7/2}$ state.
    (d) Band structures measured at 25\,K by laser ARPES with a photon energy of $h\nu=6.994$\,eV along $\bar{\mathrm{X}}$--$\bar{\Gamma}$--$\bar{\mathrm{X}}$ under $s$- (left panel) and $p$- (right panel) polarization geometries. 
    (e) Same as (d) but measured along $\bar{\mathrm{M}}$--$\bar{\Gamma}$--$\bar{\mathrm{M}}$. 
    (f) Orbital-projected band structures from nonmagnetic DFT calculations with Ce $4f$ electrons treated as core states. The dominant Ni $3d$ orbital characters are indicated by colors and symbols.
    (g) CeO-terminated slab calculations. The spectral weight projected onto the surface layers supports the assignment of the near-$\bar{\Gamma}$ branches observed in the ARPES data to surface states.}
    \label{fig:orbital_characters}
\end{figure*}

\begin{figure*}[tbp] 
    \centering
    \includegraphics[width=\textwidth]{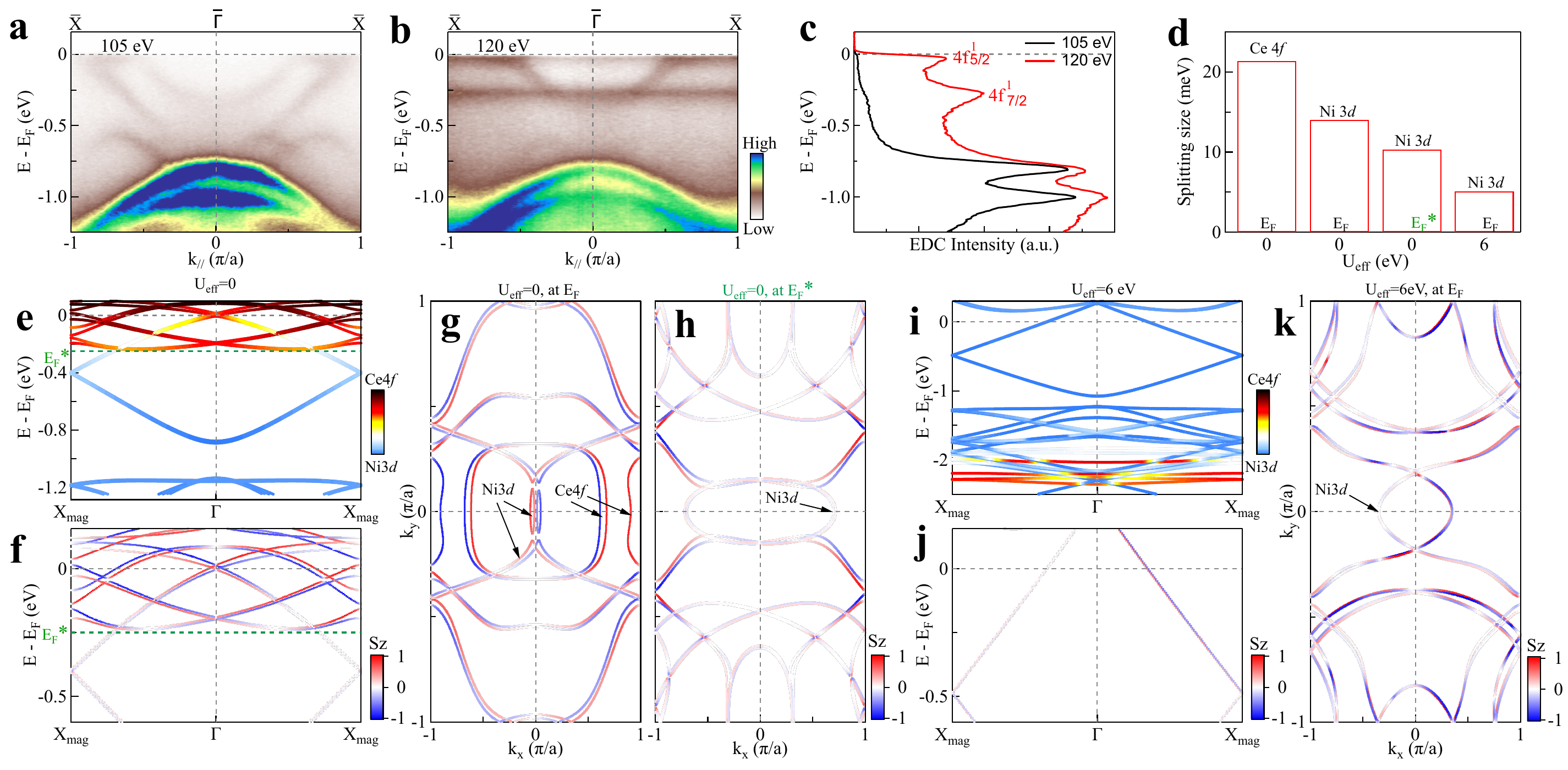}
    \caption{\textbf{Spectroscopic evidence for Ce $4f$ states and reassessment of the calculated magnetic splitting in CeNiAsO.}
    (a), (b) Band structures measured at 6\,K along $\bar{\Gamma}$--$\bar{\mathrm{X}}$ off and on the Ce $4d\rightarrow4f$ resonance, with photon energies of $h\nu=105$ and 120\,eV, respectively.
    (c) EDCs at $\bar{\Gamma}$, revealing the resonant enhancement of the $4f^1_{5/2}$ state near $E_F$ and the spin-orbit excited $4f^1_{7/2}$ state at $-0.28$\,eV.
    (d) Summary of the calculated band splitting magnitudes for four representative cases: the Ce $4f$-derived band in the itinerant-$4f$ reference, the Ni $3d$-derived band in the itinerant-$4f$ reference, the Ni $3d$-derived band after the rigid shift to $E_F^\ast$, and the Ni $3d$-derived band in the localized-$4f$ DFT+$U$ calculation. 
    (e) Orbital-projected band structures along $\mathrm{X}_{\mathrm{mag}}$--$\Gamma$--$\mathrm{X}_{\mathrm{mag}}$ from magnetic DFT calculations for the noncollinear antiferromagnetic configuration with itinerant Ce $4f$ states ($U_{\mathrm{eff}}=0$). The color scale represents the relative Ce $4f$ and Ni $3d$ orbital weights, highlighting the strong Ce $4f$ contribution near $E_F$ in the itinerant $4f$ reference. (f) Corresponding spin-projected band structures. The color scale represents the spin projection along the $z$ axis of the spin space ($S_z$). 
    (g) Corresponding calculated spin-projected Fermi surface. The Ce $4f$-derived and Ni $3d$-derived Fermi surface sheets are marked.
    (h) Same as (g) but with the Fermi level shifted to $E_F^\ast=E_F-0.25$\,eV, as indicated by green dashed lines in (e) and (f).
    (i)-(k) Same as (e)-(g) but for the localized $4f$ calculation with $U_{\mathrm{eff}}=6$\,eV. The orbital-projected bands in (i) show that the itinerant Ce $4f$ band weight is shifted away from $E_F$ in the predominantly localized $4f$ regime.
    }
    \label{fig:figure3}
\end{figure*}

\begin{figure*}[tbp] 
    \centering
    \includegraphics[width=\textwidth]{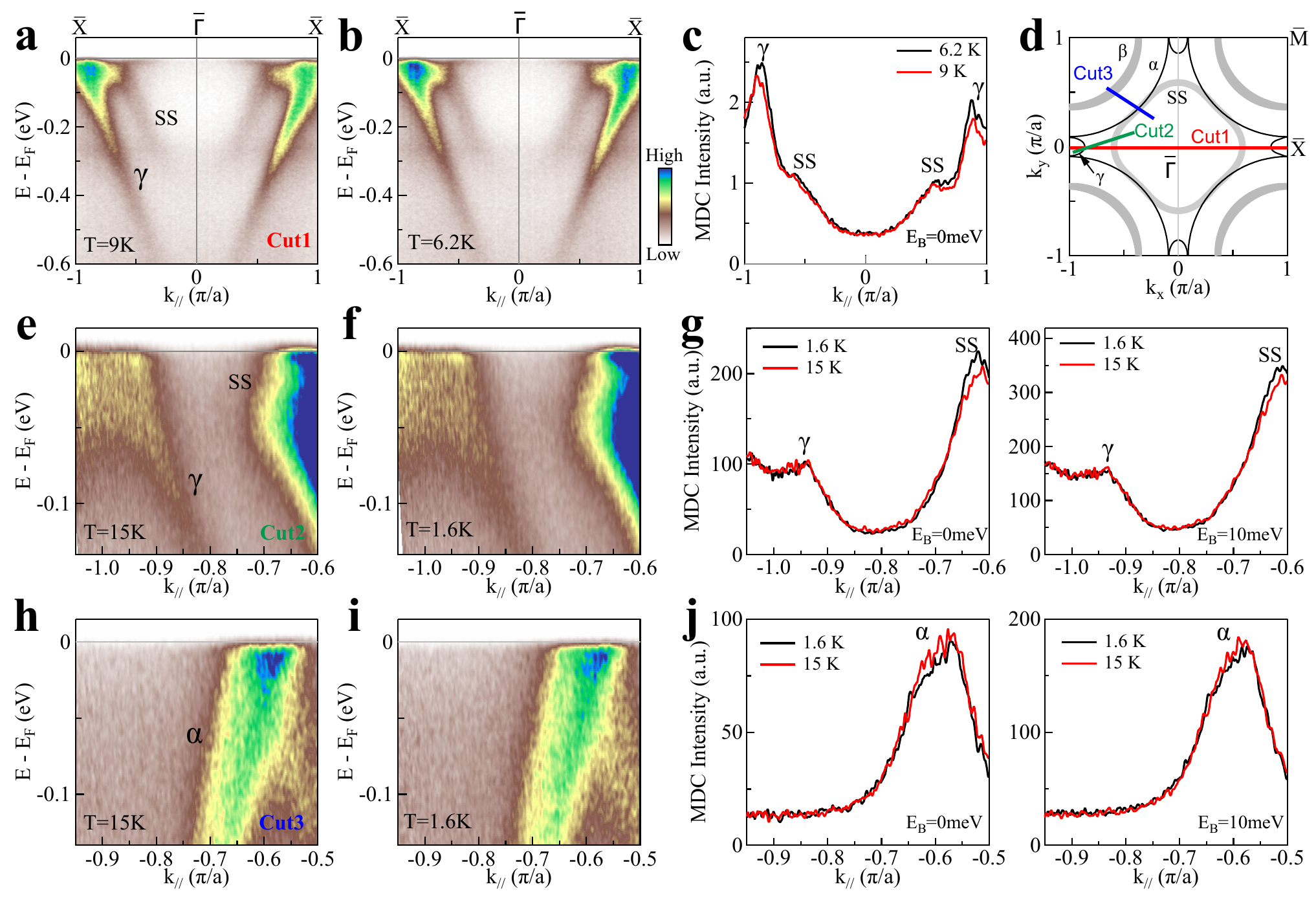}
    \caption{\textbf{Absence of resolvable $p$-wave-like exchange splitting across the successive magnetic transitions in CeNiAsO.}
    (a), (b) Band structures measured with synchrotron-based ARPES along Cut 1 at 9\,K and 6.2\,K, respectively. The momentum location of Cut 1 is defined in (d). The $\gamma$ band remains essentially unchanged across $T_{N1}\approx8.7$\,K. 
    (c) Comparison of MDCs at $E_F$ extracted from (a) and (b). The nearly overlapping MDC profiles show no resolvable exchange splitting or band folding across $T_{N1}$.
    (d) Schematic Fermi surface of CeNiAsO with the momentum locations of Cut 1, Cut 2, and Cut 3 indicated. 
    (e), (f) Band structures measured with high-resolution laser ARPES along Cut 2 at 15\,K and 1.6\,K, respectively.
    (g) MDCs at $E_B=0$ (left panel) and $E_B=10$\,meV (right panel) for 15\,K and 1.6\,K extracted from (e) and (f). 
    (h)-(j) Same as (e)-(g) but measured along the momentum Cut 3. The $\gamma$ band along Cut 2 and the $\alpha$ band along Cut 3 remain essentially unchanged from 15\,K to 1.6\,K, showing no resolvable $p$-wave-like splitting across the successive magnetic transitions at $T_{N1}$ and $T_{N2}\approx6.1$\,K.
    }
    \label{fig:figure4}
\end{figure*}

\end{document}